\input{aipcheck}

\documentclass[
    ,final            % use final for the camera ready runs
  ]
  {aipproc}

\layoutstyle{8x11double}

\begin{document}

\title{On the coupling of two quantum dots through a cavity mode}

\classification{42.50.Pq, 78.67.Hc, 03.65.Yz} 

\keywords{microcavity, quantum dots, strong coupling, emission spectrum}

\author{Elena del Valle}{
  address={School of Physics and Astronomy, University of
  Southampton, SO17 1BJ, Southampton, United Kingdom}
}

\begin{abstract}
  The effective coupling of two distant quantum dots through virtual
  photon exchange in a semiconductor microca\-vity is studied. The
  experimental conditions for strong coupling and its manifestation in
  the spectra of emission are analyzed.
\end{abstract}

\maketitle

The effective coupling of two two-level emitters through the virtual
mediation of a cavity mode has been achieved with two Rydberg
atoms~\cite{osnaghi01a} and with two superconducting
qubits~\cite{majer07a}. Very recently, hints of this coupling have
also been found between two distant quantum dots embedded in a
photonic crystal~\cite{arxiv_laucht10a,gallardo10a}. The
exper\-imental conditions for this scheme are technically demanding as
they require the simultaneous coupling of two resonant emitters to the
same cavity mode and the ability of tuning their frequency far from
it.  They, however, offer far-reaching possibilities for quantum
applications~\cite{imamoglu99a} and a laboratory for the study of the
fundamental physics of strong-coupling (SC) in a semiconductor
environment. In this text, we analyse this effective coupling between
two quantum dots (labelled $i=1,2$ with frequencies $\omega_i$),
subject to dissipation (at rates $\gamma_i$) under an incoherent
continuous excitation (at rates $P_i$) through a cavity mode (at
frequency $\omega_a$) which is not perfect (loosing photons at rate
$\gamma_a$). The two dots couple to the cav\-ity with strengths~$g_i$
and detunings $\Delta_i\equiv\omega_i-\omega_a$. We will find the
conditions and the range of parameters to achieve an effective SC
between the dots and to observe it in the spectra of emission, and
show to which extent the effective dressed states (or polaritons) are
similar to those of two dots directly coupled~\cite{delvalle10b}.

The appearance of an effective coupling between the dots occurs in the
so-called \emph{dispersive limit} which, in the absence of incoherent
processes, reduces to $\Delta\gg g$, assuming that, ideally, the dots
couple similarly to the cavity: $\Delta_1=\Delta_2=\Delta$ and
$g_1=g_2=g$. The effective coupling between the dots, $g_\mathrm{eff}=
g^2/\Delta$, is accompanied by the \emph{Stark-shift} of the bare
energies: $\omega_a'=\omega_a-2g_\mathrm{eff}$ and
$\omega_i''=\omega_i+2g_\mathrm{eff}$, for the cavity emission, and
$\omega_i'=\omega_i+g_i^2/\Delta_i$ for each dot emission. Let us fix
the units of all parameters expressing them in terms of $g$ and
consider some not very large detuning that brings us to the dispersive
limit, $\Delta=10 g$. In order to determine when the resulting
$g_\mathrm{eff}=0.1g$ dominates the dynamics, we must analyse its
effect against the incoherent processes through the spectra of
emission $S(\omega)$~\cite{delvalle_book10a}.  For instance, the
reduction of the dot-cavity couplings by detuning is affected by
decoherence as $G_i\approx
g_i/\sqrt{1+(2\Delta_i)^2/(\gamma_a+\gamma_i+P_i)^2}$~\cite{arxiv_delvalle10e}.

\begin{figure}
  \label{fig:FriJul9180311CEST2010}
  \includegraphics[width=1.85\linewidth]{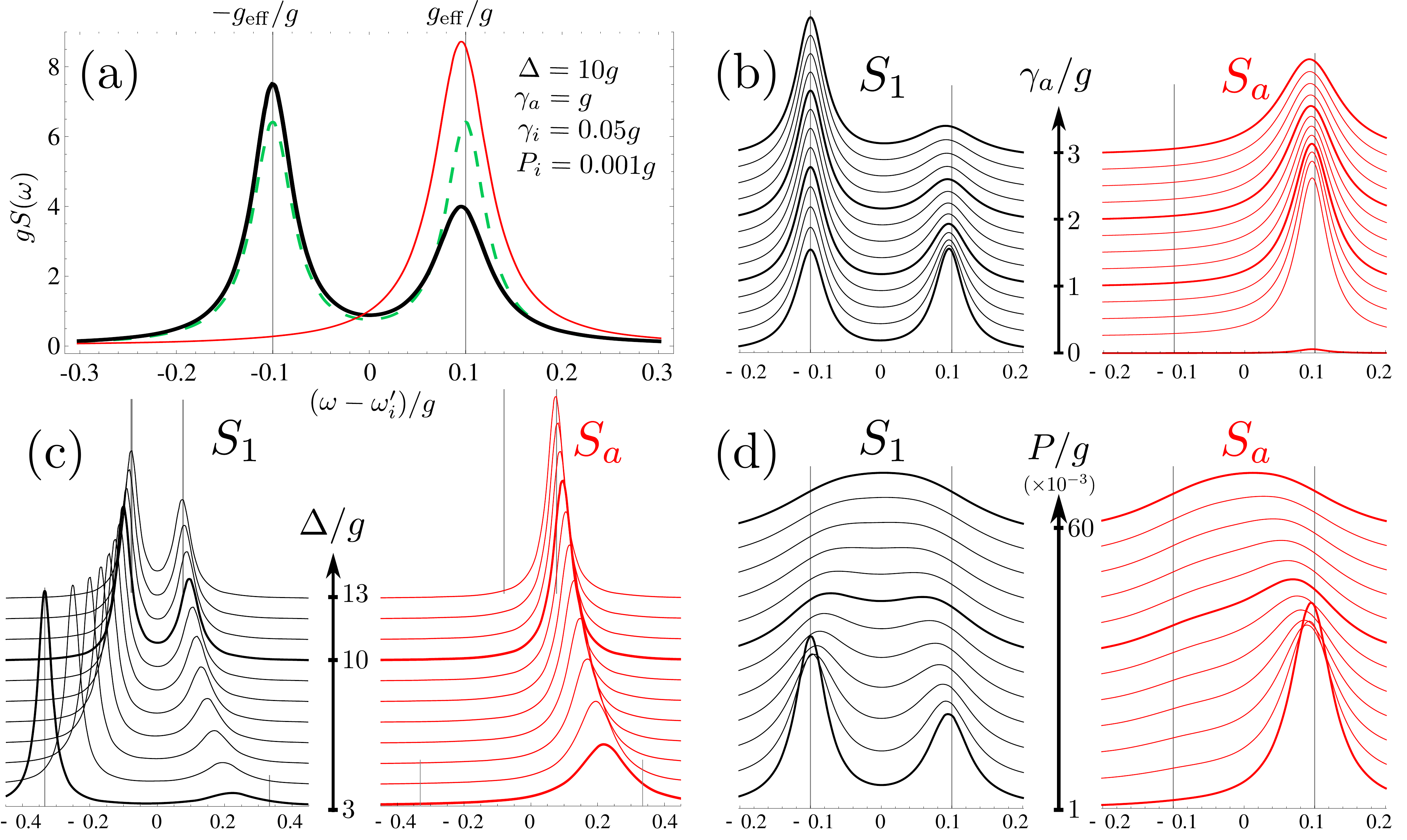}
  \caption{Normalized emission spectra of two quantum dots effectively
    coupled through a cavity: $S_1(\omega)$ (exciton spectrum, in
    thick black) and $S_a(\omega)$ (cavity spectrum, in thin red) at
    the dot frequencies $\omega_i'$ which are largely detuned from the
    cavity mode. Polariton energies ($\pm g_\mathrm{eff}$) are marked
    with vertical lines. (a) A typical experimental case (parameters
    in inset) is used to compare the effective (black) vs direct
    (dashed green) coupling and serves as a reference for panels (b-d)
    where we vary the cavity decay rate (b), the detuning from the
    cavity (c) and pumping (d), as indicated by the arrows.}
\end{figure}

In the linear regime, ($P_i\ll \gamma_i$), the system behaves almost
identically to two dots directly coupled~\cite{delvalle10b} as seen in
Fig.~\ref{fig:FriJul9180311CEST2010}(a) (dashed green). For this
figure, we have chosen reasonable parameters as compared to the state
of the art: $\gamma_i$, $P_i\ll g=\gamma_a$, where typically $g\approx
10$--$100 \,\mu eV$ and $\gamma_a\approx 20$--$200 \,\mu eV$. The dot
spectra $S_1(\omega)$ (thick black), the same for both dots unless
otherwise stated, consists of a \emph{Rabi doublet} split by
$D=2\Re\sqrt{g_\mathrm{eff}^2-[(\gamma_i+P_i)/4]^2}$ around
$\omega_i'$.  The symmetry of the Rabi doublet degrades with worsening
quality factor of the cavity, as shown in
Fig.~\ref{fig:FriJul9180311CEST2010}(b). The cavity spectra
$S_a(\omega)$ (thin red) consists of a Lorentzian at $\omega_a'$ with
linewidth given by the effective cavity rate $\Gamma_a'$\footnote{One
  can estimate the effective linewidths following
  Ref.~\cite{arxiv_delvalle10e} as: $\Gamma_a'\approx \gamma_a+\sum_i
  8G_i^2/(\gamma_i+P_i)$ and $\Gamma_i'\approx \gamma_i+P_i+
  8G_i^2/\gamma_a$.}  (not shown) plus a second Lorentzian at
$\omega_i''$ (coinciding with the upper polariton) with linewidth
given by the effective dot rate $\Gamma_i'$. This Lorentzian dominates
the cavity spectrum as $\gamma_a$ is increased. Let us note that the
cavity emission is always very weak as it is essentially empty,
$n_a\ll 1$.

One would expect that a large $\gamma_a$ (together with large
$\Delta$) is the best option to get rid of the cavity dynamics and
have the effective dot dynamics as close as possible to two directly
coupled dots, given that the condition for weak dot-cavity coupling,
$\gamma_a \gg 4 G_i $, translates approximately into
$\sqrt{\gamma_a^2+(2\Delta)^2} \gg 4 g_i$. Furthermore, a bad cavity
should not spoil the effective dot dynamics as it does involve real
photons. Therefore, we could take advantage of these premises to relax
the technical requirements on the cavity. However, $\gamma_a$ should
remain well bellow $\Delta$ so that the bare cavity and dot modes do
not overlap. This means that the purest effective dynamics (most
symmetric Rabi) is found for a perfect cavity at large $\Delta$, as
shown in Fig.~\ref{fig:FriJul9180311CEST2010}(b). The photons should
not be produced at all (be completely virtual) rather than be produced
and quickly emitted. Nevertheless, a relatively high $\gamma_a$, such
as the one in Fig.~\ref{fig:FriJul9180311CEST2010}(a), is good enough
to provide the effective coupling. Then, the approximate condition for
SC ($D>0$) between the dots is independent of $\gamma_a$. The
splitting $D$ can be enlarged by decreasing further $\Delta$, but this
has a similar side effect to increasing $\gamma_a$: it brings us out
of the dispersive limit, i. e., the symmetry of the Rabi is decreased
until the effective coupling is overcome by the direct coupling to the
cavity, as shown in
Fig.~\ref{fig:FriJul9180311CEST2010}(c). Increasing $\Delta$ too much,
on the other hand, may narrow the doublet below the detector
resolution~\cite{arxiv_laucht10a}.

Increasing pump (nonlinear regime,
Fig.~\ref{fig:FriJul9180311CEST2010}(d)), broadens and closes the Rabi
doublet, eventually bringing the system into weak coupling ($D=0$),
but it also restores the symmetry in the spectrum and converts $S_a$
into an ``echo'' of $S_1$. Effects relying on two-directly coupled
dots under incoherent excitation (extra dressed
states~\cite{delvalle10b}, entanglement in the steady
state~\cite{arxiv_delvalle10d}, etc.) can thus be approximately
reproduced by mediation of a cavity and observed through both the
cavity and exciton spectra.

Having different effective dot-cavity couplings ($G_1\neq G_2$ due to
some finite $\delta g=g_1-g_2$ and/or $\delta\Delta=\Delta_1-
\Delta_2$ with $|G_1-G_2|\ll g_\mathrm{eff}$), as it is the case
experimentally, results in a slightly different inter-dot effective
coupling but, most importantly, in different Stark shifts of the
dots. As a result, SC is probed out of resonance between the dots,
even if $\delta\Delta=0$. One can recover resonance by applying
externally the shift $\delta\Delta\approx-(g_1+g_2) \delta
g/\Delta^2$. Out of resonance, the Rabi doublet distorts into an
anticrossing whose specific shape and asymmetries depend on the
spectra, $S_1$, $S_2$ or $S_a$ (contrary to the asymmetries found
above at resonance).

A more detailed knowledge of the effect of all the parameters
(coherent and incoherent processes) on the effective coupling,
emission spectra and entanglement, can be achieved through an
effective model where the cavity degrees of freedom are included a
priori but finally traced out in the dispersive regime. This more
involved study, as well as that of the effect of pure dephasing and of
the detector resolution, is the topic of a future work.

%%%%%%%%%%%%%%%%%%%%%%%%%%%%%%%%%%%%%%%%%%%%%%%%
%% The bibliography can be prepared using the BibTeX program or
%% manually.
%%
%% The code below assumes that BibTeX is used.  If the bibliography is
%% produced without BibTeX comment out the following lines and see the
%% aipguide.pdf for further information.
%%
%% For your convenience a manually coded example is appended
%% after the \end{document}
%%%%%%%%%%%%%%%%%%%%%%%%%%%%%%%%%%%%%%%%%%%%%%%%

%%%%%%%%%%%%%%%%%%%%%%%%%%%%%%%%%%%%%%%%%%%%%%%%
%% You may have to change the BibTeX style below, depending on your
%% setup or preferences.
%%
%%
%% For The AIP proceedings layouts use either
%%%%%%%%%%%%%%%%%%%%%%%%%%%%%%%%%%%%%%%%%%%%

\bibliographystyle{aipproc}   % if natbib is available
%\bibliographystyle{aipprocl} % if natbib is missing

%%%%%%%%%%%%%%%%%%%%%%%%%%%%%%%%%%%%%%%%%%%
%% You probably want to use your own bibtex database here
%%%%%%%%%%%%%%%%%%%%%%%%%%%%%%%%%%%%%%%%%%%
\bibliography{Sci,books,qubits}

%%%%%%%%%%%%%%%%%%%%%%%%%%%%%%%%%%%%%%%%%%%
%% Just a reminder that you may have to run bibtex
%% All of it up to \end{document} can be removed
%% if you don't like the warning.
%%%%%%%%%%%%%%%%%%%%%%%%%%%%%%%%%%%%%%%%%%%
\IfFileExists{\jobname.bbl}{}
 {\typeout{}
  \typeout{******************************************}
  \typeout{** Please run "bibtex \jobname" to optain}
  \typeout{** the bibliography and then re-run LaTeX}
  \typeout{** twice to fix the references!}
  \typeout{******************************************}
  \typeout{}
 }

\end{document}